\documentstyle[12pt]{article}
\def\be{\begin{equation}}
\def\ee{\end{equation}}
\def\bea{\begin{eqnarray}}
\def\eea{\end{eqnarray}}
\begin{document}
\begin{flushright}
IMSc/04/01/01\footnote{To appear in the Proceedings of the 18th Advanced ICFA 
Beam Dynamics Workshop on Quantum Aspects of Beam Physics, October 15-20, 2000, 
Capri, Italy, Ed. Pisin Chen (World Scientific, Singapore)}  
\end{flushright}  
\begin{center} 
{\large\bf QUANTUM MECHANICS OF \\ 
DIRAC PARTICLE BEAM OPTICS: \\
\smallskip
SINGLE-PARTICLE THEORY}

\smallskip 

R. JAGANATHAN \\ 
{\it The Institute of Mathematical Sciences \\  
4th Cross Road, Central Institutes of Technology Campus \\  
Tharamani, Chennai - 600113, Tamilnadu, INDIA \\ 
E-mail: jagan@imsc.ernet.in - URL: http://www.imsc.ernet.in/$\sim$jagan}
\end{center}

\medskip

\begin{quote}
{\small It has been found that quantum corrections can substantially affect 
the classical results of tracking for trajectories close to the separatrix.  
Hence the development of a basic formalism for obtaining the quantum maps for 
any particle beam optical system is called for.  To this end, it is observed 
that several aspects of quantum maps for the beam optics of spin-$\frac{1}{2}$ 
particles can be studied, at the level of single particle dynamics, using the 
proper formalism based on the Dirac equation.}
\end{quote}

\section{Introduction}
The theory of particle beam optics, currently used in the design and operation 
of various beam devices, from electron microscopes to accelerators, is largely 
based on classical mechanics and classical electrodynamics.  Such a treatment 
has indeed been very successful in practice.  Of course, whenever it is 
essential, quantum mechanics is used in accelerator physics to understand those 
quantum effects which are prominent perturbations to the leading classical beam 
dynamics~\cite{C}.  The well-known examples are quantum excitations induced by 
synchrotron radiation in storage rings, the Sokolov-Ternov effect of spin 
polarization induced by synchrotron radiation, etc.  Recently, attention has 
been drawn by Hill~\cite{H} to the limits placed by quantum mechanics on 
achievable beam spot sizes in particle accelerators, and the need for the 
formulation of quantum beam optics relevant to such issues~\cite{V}.  In the 
context of electron microscopy scalar wave mechanics is the main tool to understand 
the image formation and its characteristics, and the spin aspects are not generally 
essential~\cite{HK}.   

In the context of accelerator physics it should be certainly desirable to have a 
unified framework based entirely on quantum mechanics to treat the orbital, 
spin, radiation, and every aspect of beam dynamics, since the constituents of 
the beams concerned are quantum particles.  First, this should help us understand 
better the classical theory of beam dynamics.  Secondly, there is already an 
indication that this is necessary too: it has been found~\cite{HY} that quantum 
corrections can substantially affect the classical results of tracking for 
trajectories close to the separatrix, leading to the suggestion that quantum maps 
can be useful in finding quickly the boundaries of nonlinear resonances.  Thus, a 
systematic formalism for obtaining the relevant quantum maps is required.  This 
problem is addressed here for the case of spin-$\frac{1}{2}$ particle beams, at the 
level of single particle dynamics as the first step towards a more comprehensive 
theory.  

\section{Quantization of the classical particle beam optics} 
If the spin is ignored, one may consider obtaining the relevant quantum maps for 
any beam optical system by quantizing the corresponding classical treatment  
directly.  The best way to do this is to use the Lie approach to classical beam 
dynamics, thoroughly developed by Dragt {\em et al.},~\cite{Det} particularly in 
the context of accelerator physics.  Ignoring the effect of spin on the orbital 
motion, the spin motion has also been treated classically, independent of the 
orbital motion, using Lie methods~\cite{Y}. 

Let the single particle optical Hamiltonian corresponding to a classical beam 
optical system be ${\cal H}(\underline{r}_\perp, \underline{p}_\perp; z)$, where 
$z$ is the coordinate along the optic axis of the system, and 
$\underline{r}_\perp = (x,y)$ and $\underline{p}_\perp = \left(p_x,p_y\right)$ 
represent the coordinates and conjugate momenta, respectively, in the transverse 
$(x,y)$-plane.  We shall assume the beam to be moving in the positive $z$-direction.  
Then for any observable of the system, 
${\cal O}(\underline{r}_\perp, \underline{p}_\perp)$, not explicitly dependent 
on $z$, the $z$-evolution equation, or the beam optical equation of motion, is 
\be 
\frac{d{\cal O}}{dz} = :-{\cal H}:{\cal O}\,,
\label{eq:eqm}
\ee
where the Lie operator $:f:$ associated with any function of the transverse 
phase-space variables, $f(\underline{r}_\perp, \underline{p}_\perp)$, is 
defined through the Poisson bracket, 
\be
:f:g = \{f,g\} 
     = \left(\frac{\partial f}{\partial x}\frac{\partial g}{\partial p_x} - 
       \frac{\partial f}{\partial p_x}\frac{\partial g}{\partial x}\right) +  
       \left(\frac{\partial f}{\partial y}\frac{\partial g}{\partial p_y} - 
       \frac{\partial f}{\partial p_y}\frac{\partial g}{\partial y}\right)\,.
\ee
When the Hamiltonian ${\cal H}$ is $z$-independent the solution of 
Eq.~(\ref{eq:eqm}) can be written down as 
\bea
{\cal O}\left(z_f\right) & = & \exp(\ell:-{\cal H}:){\cal O}\left(z_i\right) 
                               \nonumber \\ 
 & = & {\cal O}\left(z_i\right) + \ell(:-{\cal H}:{\cal O})\left(z_i\right) 
       + \left(\ell^2/2!\right)\left(:-{\cal H}:^2{\cal O}\right)\left(z_i\right) 
                               \nonumber \\ 
 & & \qquad + \left(\ell^3/3!\right)\left(:-{\cal H}:^3{\cal O}\right)
       \left(z_i\right) + \ldots \nonumber \\
 & = & {\cal O}\left(z_i\right) 
       + \ell \left(\{-{\cal H},{\cal O}\}\right)\left(z_i\right) +
     \left(\ell^2/2!\right)\left(\{-{\cal H},\{-{\cal H},{\cal O}\}\}\right)
     \left(z_i\right) \nonumber \\
 & & \qquad + 
     \left(\ell^3/3!\right)\left(\{-{\cal H},\{-{\cal H},\{-{\cal H},
     {\cal O}\}\}\}\right)\left(z_i\right) + \ldots\,, 
\label{eq:cmap}
\eea 
relating ${\cal O}\left(z_i\right)$, the value of ${\cal O}$ at an initial 
$z_i$, with ${\cal O}\left(z_f\right)$, its value at a final $z_f$, where 
$z_f > z_i$ and $\ell = \left(z_f-z_i\right)$.  When the Hamiltonian depends on 
$z$ we would have 
\be
{\cal O}\left(z_f\right) = \left(\wp\left[\exp\left(
  \int_{z_i}^{z_f} dz:-{\cal H}:\right)\right]{\cal O}\right)\left(z_i\right) 
 = \left({\cal M}\left(z_f,z_i\right){\cal O}\right)\left(z_i\right)\,, 
\label{eq:liemap}
\ee
where the transfer map, ${\cal M}\left(z_f,z_i\right)$, a Lie transformation, 
is now an $z$-ordered exponential.  

To obtain the quantum mechanical formalism for the above system we can follow 
the canonical quantization rule 
$\{\phantom{X},\phantom{Y}\} \longrightarrow 
\frac{1}{i\hbar}[\phantom{X},\phantom{Y}]$ 
where $[\phantom{X},\phantom{Y}]$ represents the commutator bracket between the 
corresponding quantum operators.  This turns Eq.~(\ref{eq:eqm}) into the 
Heisenberg equation of motion 
\be
\frac{d\hat{\cal O}}{dz} = \frac{i}{\hbar}\left[\hat{\cal H},
\hat{\cal O}\right]\,, 
\label{eq:hpeq}
\ee
where the quantum Hamiltonian operator $\hat{\cal H}$, and $\hat{\cal O}$ 
for any observable, are obtained from their respective classical counterparts 
by the replacement
\be
\underline{r}_\perp \longrightarrow \underline{\hat{r}}_\perp 
 = \underline{r}_\perp = (x,y)\,, \qquad 
\underline{p}_\perp \longrightarrow \underline{\hat{p}}_\perp = 
-i\hbar\underline{\nabla}_\perp 
= \left( -i\hbar\frac{\partial}{\partial x}, 
         -i\hbar\frac{\partial}{\partial y}\right)\,, 
\ee
followed by a symmetrization to ensure that the quantum operators are hermitian. 

From the Heisenberg picture of Eq.~(\ref{eq:hpeq}) let us go to the   
Schr\"{o}dinger picture in which a wavefunction    
$\psi\left(\underline{r}_\perp;z\right)$ is associated with the transverse  
plane at $z$.  The $z$-evolution of $|\psi(z)\rangle$ is governed by the 
beam optical Schr\"{o}dinger equation 
\be 
i\hbar\frac{\partial}{\partial z}|\psi(z)\rangle = \hat{\cal H}|\psi(z)\rangle\,.
\label{eq:speq}
\ee
Since $\left|\psi\left(\underline{r}_\perp;z\right)\right|^2$ will represent the 
probability density in the transverse plane at $z$ the average of any  
$\hat{\cal O}$ at $z$ will be 
\be
\langle\hat{\cal O}\rangle(z) = \int\int dxdy \psi^*(z)\hat{\cal O}\psi(z) 
 = \langle\psi(z)|\hat{\cal O}|\psi(z)\rangle\,,
\ee
with $\psi\left(\underline{r}_\perp;z\right)$ normalized as 
$\left\langle\psi(z)|\psi(z)\right\rangle = 1$. 

The formal solution of Eq.~(\ref{eq:speq}) is, with $|\psi_i\rangle$ $=$ 
$|\psi(z_i)\rangle$ and $|\psi_f\rangle$ $=$ $|\psi(z_f)\rangle$,  
\be
|\psi_f\rangle = \hat{U}\left(z_f,z_i\right)|\psi_i\rangle 
 = \hat{U}_{fi}|\psi_i\rangle\,, \qquad 
\hat{U}_{fi} = \wp\left[\exp\left(-\frac{i}{\hbar}\int_{z_i}^{z_f} 
   dz\hat{\cal H}\right)\right]\,.
\ee 
Thus, we get 
\be 
\langle\hat{\cal O}\rangle_f = \langle\hat{\cal O}\rangle(z_f)  
 = \langle\psi_f|\hat{\cal O}|\psi_f\rangle  
 = \langle\psi_i|\hat{U}_{fi}^\dag\hat{\cal O}\hat{U}_{fi}|\psi_i\rangle  
 = \langle\hat{U}_{fi}^\dag\hat{\cal O}\hat{U}_{fi}\rangle_i\,. 
\ee
From the correspondence between Eq.~(\ref{eq:eqm}) and Eq.~(\ref{eq:hpeq}) 
it follows immediately that 
\be
\hat{U}_{fi}^\dag\hat{\cal O}\hat{U}_{fi} 
 = \left(\wp\left[\exp\left(\int_{z_i}^{z_f}dz
   :\frac{i}{\hbar}\hat{\cal H}:\right)\right]\right)\hat{\cal O} 
 = \hat{\cal M}\left(z_f,z_i\right)\hat{\cal O}\,, 
\ee
with the definition 
$:\frac{i}{\hbar}\hat{\cal H}:\hat{\cal O}  
 = \frac{i}{\hbar}\left[\hat{\cal H},\hat{\cal O}\right]$.  
Note that in the classical limit, when 
$:\frac{i}{\hbar}\hat{\cal H}:\hat{\cal O}$ $\longrightarrow$ 
$:-{\cal H}:{\cal O}$, the quantum Lie transformation 
$\hat{\cal M}\left(z_f,z_i\right)$ becomes the classical Lie transformation 
${\cal M}\left(z_f,z_i\right)$.  This shows that if a system corresponds  
classically to a map 
\be
(\underline{r}_{\perp i},\underline{p}_{\perp i}) \longrightarrow 
(\underline{r}_{\perp f},\underline{p}_{\perp f}) 
 = (\underline{R}_\perp(\underline{r}_{\perp i}, \underline{p}_{\perp i}),
    \underline{P}_\perp(\underline{r}_{\perp i}, \underline{p}_{\perp i}))\,,  
\ee 
then it will correspond to a map of quantum averages as given by 
\be 
\langle\underline{\hat{r}}_\perp\rangle_i \longrightarrow  
\langle\underline{\hat{r}}_\perp\rangle_f = 
\langle\underline{\hat{R}}_\perp(\underline{\hat{r}}_\perp,
       \underline{\hat{p}}_\perp)\rangle_i\,, \quad  
\langle\underline{\hat{p}}_\perp\rangle_i \longrightarrow   
\langle\underline{\hat{p}}_\perp\rangle_f = 
\langle\underline{\hat{P}}_\perp(\underline{\hat{r}}_\perp,
       \underline{\hat{p}}_\perp)\rangle_i\,. 
\label{eq:qmap}
\ee 

To see what Eq.~(\ref{eq:qmap}) implies let us consider, for example, a 
classical Lie transformation $\exp\left(:\frac{a}{3}x^3:\right)$ corresponding 
to a kick in the $xz$-plane by a thin sextupole.  This leads to the classical 
phase-space map 
\be
x_f = x_i\,, \qquad p_f = p_i + a x_i^2\,,
\ee
as follows from Eq.~(\ref{eq:liemap}).  This would correspond to the quantum Lie 
transformation $\exp(:\frac{a}{3}\hat{x}^3:)$ which leads, as seen from 
Eq.~(\ref{eq:qmap}), to the following map for the quantum averages: 
\be
\langle\hat{x}\rangle_f = \langle\hat{x}\rangle_i\,, \quad 
\langle\hat{p}\rangle_f = \langle\hat{p}\rangle_i + a\langle\hat{x}^2\rangle_i 
 = \langle\hat{p}\rangle_i + a\langle\hat{x}\rangle_i^2 
  + a\langle(\hat{x}-\langle\hat{x}\rangle)^2\rangle_i\,.  
\label{eq:xpqmap}
\ee
Now, we can consider the expectation values, such as $\langle\hat{x}\rangle$ 
and $\langle\hat{p}\rangle$, as corresponding to their classical values 
{\it \`{a} la} Ehrenfest.  Then, as the above simple example shows, generally, 
the leading quantum effects on the classical beam optics can be expected to be 
due to the uncertainties in the initial conditions like the term   
$a\langle(\hat{x}-\langle\hat{x}\rangle)^2\rangle_i$ in Eq.~(\ref{eq:xpqmap}).  
As pointed out by Heifets and Yan,~\cite{HY} such leading quantum corrections 
involve the Planck constant $\hbar$ not explicitly but only through the uncertainty 
principle which controls the minimum limits for the initial conditions.  This has 
been realized earlier also~\cite{Jet1,KTh,Jet2}, particularly in the context of 
electron microscopy~\cite{Jet1,KTh}.  In a detailed study~\cite{HY} of a simple 
example it has been found that trajectories close to the separatrix are strongly 
perturbed in spite of very small initial rms ($10^{-15}$) and small ($1500$) number 
of turns. 

As is clear from the above, a quantum formalism derived from the classical beam 
optics can be expected to give all the leading quantum corrections to the classical 
maps.  The question that arises is how to go beyond and obtain the quantum maps 
more completely starting {\em ab initio} with the quantum mechanics of the 
concerned system since such a process should lead to other quantum corrections not 
derivable simply from the quantization of the classical optical Hamiltonian.  
Essentially, one should obtain the quantum beam optical Hamiltonian $\hat{\cal H}$ 
of Eq.~(\ref{eq:speq}) directly from the original time-dependent Schr\"{o}dinger 
equation of the system.  Once $\hat{\cal H}$ is obtained Lie methods~\cite{Det,Y} 
can be used to construct the quantum $z$-evolution operator $\hat{U}_{fi}$ and study 
the consequent quantum maps.  Derivations of $\hat{\cal H}$ for the Klein-Gordon 
and Dirac particle beams will be discussed in the following sections.   
  
A more complete theory, even at the level of optics, must take into account 
multiparticle effects.  To this end, it might be profitable to be guided by the 
models developed by Fedele {\em et al.}~\cite{Fet,Ket} (thermal wave model - TWM) 
and Cufaro Petroni {\em et al.}~\cite{Cet} (stochastic collective dynamical 
model - SCDM) for treating the beam phenomenologically as a quasiclassical many-body 
system.  Though the details of approach and interpretation are different, both 
these models suggest phenomenological Schr\"{o}dinger-like wavefunction descriptions 
for the collective motion of the beam.  In TWM the beam emittance plays the role of 
$\hbar$.  In SCDM it is argued that $\hbar$ is to be replaced by an effective unit 
of beam emittance given in terms of the Compton wavelength of the beam particle and 
the number of particles in the beam.  It may be noted that Lie algebraic tools can 
be used to handle any Schr\"{o}dinger-like equation. 
   
\section{Using the Klein-Gordon equation ignoring the spin} 
One may consider getting a theory of quantum maps for spin-$\frac{1}{2}$ 
particle beam optical system based on the Klein-Gordon equation ignoring the 
spin.  For this, one has to transform the equation 
\be
\left(i\hbar\frac{\partial}{\partial t}-q\hat{\phi}\right)^2
 \Psi\left(\underline{r}_\perp, z;t\right) = 
\left\{c^2\left[\underline{\hat{\pi}}_\perp^2 
 + \left(-i\hbar\frac{\partial}{\partial z}-q\hat{A}_z\right)^2\right] 
 + m^2c^4\right\}\Psi\left(\underline{r}_\perp, z;t\right)\,,
\label{eq:kgeq}
\ee 
into the beam optical form in Eq.~(\ref{eq:speq}); in Eq.(\ref{eq:kgeq}) 
$q$ is the charge of the particle, $\hat{\pi}_\perp$ $=$ 
$(\hat{\pi}_x,\hat{\pi}_y)$ $=$ $(\hat{p}_x-q\hat{A}_x , \hat{p}_y-q\hat{A}_y)$, 
$\underline{\hat{\pi}}_\perp^2$ $=$ $\hat{\pi}_x^2 + \hat{\pi}_y^2$, and $\phi$ and 
$\underline{A} = \left(A_x,A_y,A_z\right)$ are, respectively, the scalar  
and vector potentials of the electric and magnetic fields of the optical system 
($\underline{E} = -\underline{\nabla}\phi$, 
$\underline{B} = \underline{\nabla}\times\underline{A}$).  In the standard 
relativistic quantum theory~\cite{BD}, Feshbach-Villars and Foldy-Wouthuysen 
techniques are used for reducing the Klein-Gordon equation to its nonrelativistic 
approximation plus the relativistic corrections.   Applying analogous techniques in 
the special case of a quasiparaxial ($|\underline{p}_\perp| \ll p_z$) monoenergetic 
beam propagating through a system with time-independent fields one can reduce 
Eq.~(\ref{eq:kgeq}) to the beam optical form of Eq.~(\ref{eq:speq}) with 
$\hat{\cal H}$ containing a leading paraxial part followed by nonparaxial 
parts~\cite{Jet1,KTh,Jet3}.  In this case the wavefunction in Eq.~(\ref{eq:kgeq}) 
can be assumed to be of the form 
\be
 \Psi(\underline{r}_\perp, z;t) = \psi(\underline{r}_\perp; z)
 \exp\left[\frac{i}{\hbar}\left(p_0z-Et\right)\right]\,, 
\label{eq:psi}
\ee
where $p_0$ is the design momentum of the beam and $E = +\sqrt{c^2p_0^2+m^2c^4}$. 
Then the resulting time-independent equation for $\psi(\underline{r}_\perp;z)$ can 
be regarded as describing the scattering of the beam particle by the system and  
transformed into an equation of the type in Eq.~(\ref{eq:speq})~\cite{Jet1,KTh,Jet3}.   

For example, for a normal magnetic quadrupole lens with 
$\underline{A} = (0,0,\frac{1}{2}K(x^2-y^2))$, where $K$ is nonzero inside the 
lens region and zero outside, 
\be
\hat{\cal H} \approx \frac{1}{2p_0}\left(\hat{p}_x^2+\hat{p}_y^2\right) 
               -\frac{1}{2}qK\left(\hat{x}^2-\hat{y}^2\right) 
               +\frac{1}{8p_0^3}\left(\hat{p}_x^2+\hat{p}_y^2\right)^2 
               +\frac{qK\hbar^2}{4p_0^4}\left(\hat{p}_x^2-\hat{p}_y^2\right)\,.  
\label{eq:kgoh} 
\ee
It must be noted that while the first three terms of $\hat{\cal H}$ in 
Eq.~(\ref{eq:kgoh}) are exactly the terms derivable by direct quantization of the 
classical beam optical Hamiltonian the last, $\hbar$-dependent, term is a quantum 
correction not derivable from the classical theory.  Though such $\hbar$-dependent 
terms may seem to be too small, particularly for high energy beams, they may become 
effective when there are large fluctuations in the initial conditions since they 
essentially modify the coefficients in the classical maps.  

\section{The proper theory using the Dirac equation}
For a spin-$\frac{1}{2}$ particle beam the proper theory should be based on the 
Dirac equation if one wants to treat all the aspects of beam optics including 
spin evolution and spin-orbit interaction.  In such a case the Schr\"{o}dinger 
equation to start with is 
\be
i\hbar\frac{d}{dt}\Psi\left(\underline{r}_\perp, z;t\right) 
  = \hat{H}\Psi\left(\underline{r}_\perp, z;t\right)\,, 
\ee
where $\Psi$ is now a $4$-component spinor and $\hat{H}$ is the Dirac Hamiltonian 
\be
\hat{H} = \beta 
mc^2+q\hat{\phi}+c\underline{\alpha}_\perp\cdot\underline{\hat{\pi}}_\perp
          + c\alpha_z\left(-i\hbar\frac{\partial}{\partial z}-q\hat{A}_z\right) 
          -\mu_a\beta\underline{\Sigma}\cdot\underline{B}\,,
\label{eq:dh}
\ee
including the Pauli term to take into account the anomalous magnetic moment 
$\mu_a$.  In Eq.~(\ref{eq:dh}) all the symbols have the usual meanings as in the 
standard Dirac theory~\cite{BD}.  Considering the special case of a quasiparaxial 
monoenergetic beam we can take $\Psi\left(\underline{r}_\perp, z;t\right)$ to be 
of the form in Eq.~(\ref{eq:psi}).  Then the $4$-component  
$\psi(\underline{r}_\perp; z)$ satisfies the time-independent Dirac equation 
\be
\left[\beta mc^2+q\hat{\phi}
 +c\underline{\alpha}_\perp\cdot\underline{\hat{\pi}}_\perp
 + c\alpha_z\left(-i\hbar\frac{\partial}{\partial z}-q\hat{A}_z\right) 
          -\mu_a\beta\underline{\Sigma}\cdot\underline{B}\right]
          \psi\left(\underline{r}_\perp; z\right) 
 = E\psi\left(\underline{r}_\perp; z\right)\,. 
\label{eq:tide}
\ee
describing the scattering of the beam particle by the system.  

Actually Eq.~(\ref{eq:tide}) has the ideal structure for our purpose since it is 
already linear in $\frac{\partial}{\partial z}$.  So one can readily rearrange the 
terms in it to get the desired form of Eq.~(\ref{eq:speq}).  However, it is 
difficult to work directly with such an equation since there are problems 
associated with the interpretation of the results using the traditional 
Schr\"{o}dinger position operator~\cite{T}.  In the standard theory the 
Foldy-Wouthuysen (FW) transformation technique is used to reduce the Dirac 
Hamiltonian to a form suitable for direct interpretation in terms of the 
nonrelativistic part and a series of relativistic corrections.  Derbenev and 
Kondratenko (DK)~\cite{DK} used the FW technique to get their Hamiltonian for 
radiation calculations.  Heinemann and Barber~\cite{HB} have reviewed the 
derivation of the DK Hamiltonian and have used it to suggest a quantum formulation 
of Dirac particle beam physics, particularly for polarized beams, in terms of machine 
coordinates, observables, and the Wigner function.  

In an independent and different approach an FW-like technique has been used to 
develop a systematic formalism of Dirac particle beam optics in which the aim 
has been to expand the Dirac Hamiltonian as a series of paraxial and nonparaxial 
approximations~\cite{Jet1,KTh,Jet2,Jet4,CJKP}.  This leads to the reduction of 
the original $4$-component Dirac spinor to an effective $2$-component  
$\psi(\underline{r}_\perp; z)$ which satisfies the accelerator optical 
Schr\"{o}dinger equation~\cite{CJKP} 
\be
i\hbar\frac{\partial}{\partial z}\psi\left(\underline{r}_\perp; z\right) 
= \hat{\cal H}\psi\left(\underline{r}_\perp; z\right)\,, \qquad
\psi\left(\underline{r}_\perp; z\right) = 
\left(\begin{array}{c}
\psi_1\left(\underline{r}_\perp; z\right) \\
\psi_2\left(\underline{r}_\perp; z\right)
      \end{array}\right)\,, 
\label{eq:aose} 
\ee
where $\hat{\cal H}$ is a $2 \times 2$ matrix operator incorporating the 
Stern-Gerlach (SG) spin-orbit effect and the Thomas-Bargmann-Michel-Telegdi (TBMT) 
spin evolution.  As is usual in accelerator theory the spin operator 
$\underline{S} = \frac{1}{2}\hbar\underline{\sigma}$ entering the accelerator 
optical Hamiltonian $\hat{\cal H}$ refers to the rest frame of the moving particle.  
Further, $(\hat{x}, \hat{y})$ and $(\hat{p}_x, \hat{p}_y)$ in $\hat{\cal H}$ 
correspond to the observed particle position and momentum components in the 
transverse plane.  It should be noted that the $2$-component 
$\psi(\underline{r}_\perp; z)$ of Eq.~(\ref{eq:aose}) is an accelerator optical 
approximation of the original $4$-component Dirac spinor, valid for any value of 
the design momentum $p_0$ from nonrelativistic to extreme relativistic region.  
 
For the normal magnetic quadrupole lens the accelerator optical Hamiltonian reads
\bea 
\hat{\cal H} & \approx & \frac{1}{2p_0}\left(\hat{p}_x^2+\hat{p}_y^2\right) 
            -\frac{1}{2}qK\left(\hat{x}^2-\hat{y}^2\right) 
            +\frac{1}{8p_0^3}\left(\hat{p}_x^2+\hat{p}_y^2\right)^2 
            +\frac{q^2K^2\hbar^2}{8p_0^3}\left(\hat{x}^2+\hat{y}^2\right) 
\nonumber \\ 
  &   & \qquad +\frac{(q+\gamma \epsilon)K}{p_0}\left(\hat{x}S_y+\hat{y}S_x\right)\,, 
\label{eq:dqh} 
\eea
where $\gamma = E/mc^2$ and $\epsilon = 2m\mu_a/\hbar$.  The last spin-dependent 
term accounts for the SG kicks in the transverse phase-space and the TBMT spin 
evolution.  As in the Klein-Gordon case of Eq.~(\ref{eq:kgoh}), $\hat{\cal H}$ of 
Eq.~(\ref{eq:dqh}) also contains all the terms derivable from the classical theory 
plus the quantum correction terms.  But, it must be noted that the scalar quantum 
correction term in Eq.~(\ref{eq:dqh})) (4th term) is not the same as the 4th term 
in Eq.~(\ref{eq:kgoh}).  Thus, besides in the $\hbar$-dependent effects of spin 
on the orbital quantum map ({\em e.g.}, the last term in Eq.~(\ref{eq:dqh})), even 
in the $\hbar$-dependent scalar quantum corrections the Dirac particle has its own 
signature different from that of the Klein-Gordon particle.  

\section{Conclusion}
The problem of obtaining the quantum maps for phase-space transfer across particle 
beam optical systems has been reviewed.  The leading quantum corrections to the 
classical maps are mainly due to the initial uncertainties and involve the Planck 
constant $\hbar$ not explicitly but only through the minimum limits set by the 
uncertainty principle.  These corrections can be obtained by direct quantization 
of the Lie algebraic formalism of classical particle beam optics.  The Klein-Gordon 
and Dirac theories add further subtle, $\hbar$-dependent, corrections which may 
become effective when there are large fluctuations in the initial uncertainties.  
Contrary to the common expectation the 
scalar approximation of the Dirac theory is not completely equivalent to the 
Klein-Gordon theory.  All aspects of quantum maps for spin-$\frac{1}{2}$ particle 
beams, including spin evolution and spin-orbit effects, can be studied, at the level 
of single particle dynamics, using the proper formalism based on the Dirac equation. 

\section*{Acknowledgements} 
I am grateful to M. Pusterla and R. Fedele who made it possible for me to 
participate in the QABP2K Workshop through financial support, by INFN-Napoli 
and INFN-Padova, for international travel, and accommodation expenses at Napoli and 
Capri.  My special thanks are due to R. Fedele for the kind hospitality and 
stimulating discussions during my visit to the INFN-Napoli.  I am thankful to S. De 
Siena, S. De Martino, and F. Illuminati for the warm hospitality and useful 
discussions during my visit to the Department of Physics, University of Salerno, 
during the week of the Mini Workshop on Quantum Methodologies in Beam Physics.   


\end{document}